# Diffusion NMR Methods Applied to Xenon Gas for Materials Study

R. W. Mair[1,2], M. S. Rosen[1], R. Wang[2], D. G. Cory[2], and R. L. Walsworth[1]

[1] Harvard-Smithsonian Center for Astrophysics, Cambridge, MA 02138, USA

[2] Dept. of Nuclear Engineering, Massachusetts Institute of Technology, Cambridge, MA 02139, USA

**Corresponding Author:**

Ross Mair

Harvard Smithsonian Center for Astrophysics,

60 Garden St, MS 59,

Cambridge, MA, 02138,

USA

Phone:  1-617-495 7218

Fax:  1-617-496 7690

Email: rmair@cfa.harvard.edu

# ABSTRACT

We report initial NMR studies of i) xenon gas diffusion in model heterogeneous porous media, and ii) continuous flow laser-polarized xenon gas. Both areas utilize the Pulsed Gradient Spin Echo techniques in the gas-phase, with the aim of obtaining more sophisticated information than just translational self-diffusion coefficients - a brief overview of this area is provided in the introduction. The heterogeneous or multiple-length scale model porous media consisted of random packs of mixed glass beads of two different sizes. We focus on observing the approach of the time-dependent gas diffusion coefficient, $D(t)$, (an indicator of mean squared displacement) to the long-time asymptote, with the aim of understanding the long-length scale structural information that may be derived from a heterogeneous porous system. We find $D(t)$ of imbibed xenon gas at short diffusion times is similar for the mixed bead pack and a pack of the smaller sized beads alone, hence reflecting the pore surface-area-to-volume-ratio of the smaller bead sample. The approach of $D(t)$ to the long-time limit follows that of a pack of the larger sized beads alone, although the limiting $D(t)$ for the mixed bead pack is lower, reflecting the lower porosity of the sample compared to that of a pack of mono-sized glass beads. The Padé approximation is used to interpolate $D(t)$ data between the short and long time limits. Initial studies of continuous flow laser-polarized xenon gas demonstrate velocity-sensitive imaging of much higher flows than can generally be obtained with liquids (20 - 200 mm/s). Gas velocity imaging is, however, found to be limited to a resolution of about 1 mm/s due to the high diffusivity of gases compared to liquids. We also present the first gas-phase NMR scattering, or diffusive-diffraction, data: namely, flow-enhanced structural features in the echo attenuation data from laser-polarized xenon flowing through a 2 mm glass bead pack.

# INTRODUCTION

*Overview - PGSE Methods*

NMR methods for measuring translational molecular diffusion (i.e., Brownian motion), based on the Pulsed Gradient Spin Echo (PGSE) technique, have, over the last 15 years, become widely accepted practice in just about every field in which NMR or MRI is used. It was the pioneers of NMR who very early recognized that the NMR signal in a spin-echo measurement could be susceptible to molecular diffusion effects [1]. Stejskal and Tanner introduced the PGSE method in 1965 [2] and demonstrated its power to measure translational molecular diffusion of liquids or solutes in a liquid phase. In the last 10 years, with advances in hardware on commercial vertical-bore NMR spectrometers (to the point where pulsed field gradient capability is now virtually standard), PGSE-based diffusion methods have become a powerful tool in physical and biophysical chemistry. These methods permit accurate measures of self-diffusion coefficients [3], aid studies of molecular association and phase separation, [3,4] macromolecular tumbling [5], and multiple-component mixture analysis [6,7].

The original pioneers of PGSE also noted that the diffusion measurement, when applied to heterogeneous systems, could be influenced by details of the microstructure of the system [8]. The diffusion propagator [8,9] and q-space imaging [10,11] formalisms for interpreting PGSE experiments recognized that the echo attenuation data was the Fourier conjugate of the enclosing medium's structure factor. This provided a new level of access to microstructural information of porous or enclosing media at length scales more than an order of magnitude less than what can be resolved with conventional NMR imaging (MRI) [12]. A theoretical understanding of the propagator formalism, with respect to the influence of the time over which diffusion in porous media is measured [13], proved that additional structural information relating to pore surface-area-to-volume-ratio [14,15] and long-length scale pore connectivity (tortuosity) [16,17] could also be obtained from PGSE experiments. Such methods have more recently been applied in biophysics research, providing measures of cell dimensions [18] and nerve fiber compartmentation [19].

The same pulse sequences that yield self diffusion coefficients or microstructural detail of enclosing media via restricted diffusion can also provide accurate measurements of coherent velocity, by determining the offset of the diffusion propagator from zero [12,20]. These experiments are most powerful when combined with traditional NMR imaging, to produce images of samples experiencing flow, with pixel intensities indicative of regional velocity. Such methods have been used in rheological investigations of complex fluids [20-22], structural characterization of porous media, where the flow enhances structural feature factors [23-25], and studies of convection and non-uniform flow in liquid systems [26,27]. In clinical medicine, the combination of MRI and diffusion coefficient measurement (or diffusion weighting) is a powerful indicator of the onset of stroke [28]. Anisotropic diffusion, described by a three-dimensional tensor [29] can be used to indicate neural fiber orientation [30], and permit tracking of neural fibers through the body [31].

A complete review of all the above fields, and the application of the NMR diffusion measurement (PGSE technique) within them would be extensive, and is beyond the scope of this article. However, from this brief introduction, it is clear the PGSE technique is a powerful NMR method that has touched nearly every field that uses NMR or MRI.

*Overview - Gas Phase PGSE NMR*

Until a decade ago, gas-phase NMR had been a very limited field of study. Generally, this was a result of the very low spin density of NMR-sensitive gases, due to the ~ 1000 times lower density than any liquid phase. The most common applications had been the study of molecular dynamics throughout the phase diagram of fluorinated or noble gases via NMR relaxation times [32,33] and the use of fluorinated gases to image porous and biological samples [34-36]. In the last 6-8 years, however, there has been a surge of interest in gas-phase NMR, following the application of optical pumping techniques using lasers to produce large nuclear spin polarizations ($\gtrsim$ 10%) in the spin 1/2 noble gases ($^{129}$Xe and $^3$He) [37]. Most initial interest in laser polarized noble gas NMR focused on imaging of the lung gas space in animals [38] and humans [39]. This dominant trend has continued, and recent reviews have summarized the extensive progress in these fields, especially with respect to human imaging [40-42].

However, gas-phase NMR, either laser- or thermally-polarized is also a very useful tool for applications other than lung imaging. We have recently summarized these applications [43]. With very little experimental interest in gas-phase NMR before 1994, there had also been only sporadic attempts to apply traditional NMR diffusion techniques to such systems, measuring the pressure or gas-mixture-dependence of the self-diffusion coefficient [44-46]. With the advent of laser polarized noble gas NMR, several publications addressed the measurement of noble gas diffusion coefficients while accounting for the non-renewable magnetization provided by the laser polarization technique. Initially, a simple one-dimensional imaging technique was used, bleaching a hole in the gas magnetization and watching the signal return over time [47-49]. More recently [49-52], a method based on the original technique of Carr and Purcell [1] for measuring unrestricted diffusion in a constant magnetic field gradient using a spin-echo has been employed. However, rather than using a 180° RF pulse to form a spin-echo, the sign of the gradient was reversed midway through the measurement period and the resulting gradient echo performed the refocusing.

We began looking at more sophisticated applications of the PGSE-related techniques, as described in the previous section, and the advantages provided by studying gases with such techniques. We addressed specific problems relating to the implementation, and limitations, of these techniques with laser-polarized gases [53,54]. Specifically, as the magnetization in a sample of laser-polarized gas arises from the optical-pumping process outside the magnet [37], this magnetization is independent of the magnetic field. Furthermore, once consumed by RF pulses, the magnetization will not return after an appropriate relaxation delay - equilibrium magnetization is restored instead, which is 4 - 5 orders of magnitude smaller than the laser-polarized magnetization. Therefore, efficient use of the magnetization is required, which can be accomplished in one of three ways. i) Using small flip-angle excitation pulses can provide multiple excitations, however they can generally only be used with single-pulse methods such as gradient echo experiments. Additional 180° RF pulses suffer from $B_1$ inhomogeneity and pulse-miscalibration and can provide as much spurious signal as the small flip-angle excitation pulse, while successive 90° pulses in stimulated echoes assume no longitudinal magnetization after the excitation pulse. Furthermore, long $T_2^*$ is a requirement. ii) Constant refocusing techniques can use a 90° excitation pulse, permitting the use of multiple-pulse sequences and allowing multiple diffusion encodings during a CPMG-style echo train. However, such methods require long $T_2$, which is often not the case in heterogeneous samples. iii) The gas can be constantly replenished, either in a batch or continuous flow mode, and then the methods implemented in a traditional manner.

While novel pulse methods work successfully for measuring homogeneous gas diffusion [53,54], we settled on method iii), constantly replenishing the gas, to implement the time-dependent diffusion technique previously applied to liquid-saturated systems [15,17], in gas-saturated porous media [55]. However, without automation, the method was tedious, and we found that moderate signal-averaging with higher pressures of thermally polarized gas provided the same results [55]. Using $^{129}$Xe gas as the observation spin, with a diffusion coefficient ~ 3 orders of magnitude higher than that of water (5.7 × $10^{-6}$ m$^2$ s$^{-1}$ at 1 bar pressure [53]), it was possible to extend the porous length scales over which diffusion was observed by over an order of magnitude. We have since shown that pore length scales on the order of millimeters, rather than tens of micrometers, can be probed; and that NMR gas diffusion can be used for tortuosity determination in reservoir rocks [55-57].

In recent years, the PGSE techniques have also been exploited for studying gas flow. Two-dimensional velocity imaging of laser-polarized xenon has highlighted the convective motion of xenon gas in a two-phase gas-liquid system [58]. Additionally, laser-polarized xenon flowing in a closed loop system has been observed with one-dimensional velocity imaging in tube flow in a variety of diameters, and with obstructions [59,60]. Thermally-polarized hydrocarbon gases flowing in catalyst samples have also been studied recently [61,62]. Additionally, in the clinical applications, diffusion imaging of inhaled laser-polarized $^3$He in the lungs has become popular [63,64]. Significant changes are observed between healthy and diseased subjects [65], and correlations of the observed diffusion coefficient with other traditional lung spirometry measurements show promise [66]. However, actual determinations of structural information of the enclosing airways should be treated with caution, due to the general inability of clinical MR systems to permit adequate background gradient compensation, as has been crucial to interpreting NMR diffusion data in porous media [53,67,68].

One other area of gas-phase NMR diffusion study should be noted. It has long been known that the xenon atom's sensitivity to chemical shielding gives it a wide chemical shift range in nano-porous materials or when dissolved in organic compounds or biological tissue [69,70]. This area of study has also recently seen the introduction of PGSE techniques to studies of xenon dissolved in liquid crystals [71,72], polymers [73,74] and aqueous solution [75], and in nano-porous solids such as zeolites [76].

The above is a very brief overview of PGSE applications in gas-phase NMR. While gas-phase NMR was bought to our attention due to the advances in laser-polarization processes, this development has also prompted a realization of the usefulness of thermally polarized gases for materials and bio-medical study. Whether the gas is laser or thermally polarized, similar information can often be derived from NMR experiments, although the technique and time-scale for obtaining the data may differ. As such, we believe both laser-polarized and thermally polarized gases offer the potential for complementary information from different types of samples in porous media and other fields of materials study.

*Introduction to this work*

In this paper, we report initial experiments in two different areas of porous media study using laser and thermally polarized xenon gas. Thermally polarized xenon restricted diffusion is measured as a probe of mean squared displacement and hence long-length scale connectivity in mixed-size glass bead packs. This work continues and extends our prior studies with thermally polarized $^{129}$Xe in rocks and mono-sized glass bead packs [55-57]. In addition, we have developed, partially implemented, and begun testing a new polarized gas delivery system. In the current implementation, the system allows for continuous, accurately controlled flow of polarized xenon through a sample in the NMR magnet. We show initial tests of gas flow in tubes and PGSE measurements of flowing gas in model porous media.

In porous media study, some parameters, such as the pore surface-area-to-volume ratio can be easily measured, including by liquid-phase NMR measurements of the effective or time-dependent diffusion coefficient, $D(t)$ [14,15]. The tortuosity, which describes long-length scale pore connectivity and fluid transport properties is much more difficult to determine. While a theoretical understanding shows that this can be measured from the long-time asymptotic limit approached by $D(t)$ [16], water spins are unable to probe (diffuse) sufficient length scales in these samples during a feasible NMR measurement time, in order to reflect the true heterogeneity of the sample [17]. Our recent experiments with gas-phase NMR measurements of the $^{129}$Xe $D(t)$ (Gas Diffusion NMR) have provided the first such measurements of tortuosity in rocks and glass beads [55,56], enabled by the much longer diffusion distances experienced by gases with their diffusion coefficients ~ 3 orders of magnitude higher than water. Gas diffusion NMR also provides the only measure of the heterogeneity length of the sample, i.e., the diffusion distance at which the true tortuosity is experienced [56,57]. While experiments with mono-sized glass beads show a simple relationship of $D(t)$ with diffusion length [57], experiments with rocks showed non-monotonic changes in $D(t)$, indicative of multiple length scales in the sample [55]. Gas diffusion NMR is the only method that permits studies of the approach of $D(t)$ to the tortuosity limit, and hence has the potential to provide information about the multiple length scales in heterogeneous systems. In this paper, we report initial measurements of $^{129}$Xe $D(t)$ in model heterogeneous porous systems - i.e., samples of mixed glass bead packs, with two different sized glass beads, as the beginning of a study to try to obtain greater insight into the length-scale information that may be derived from the approach of $D(t)$ to the asymptotic limit.

In addition to experiments with static fluids, fluid flows can also be used to study microstructural detail in porous media [23-25], as well as other transport phenomena in complex flows [27-28]. Many of these experiments have not yet exploited gas-phase NMR, and in general, such phenomena will only be observable in a feasible laboratory timescale using laser-polarized gas. In addition to obtaining faster gas-phase $D(t)$ measurements, and probing peculiar characteristics of gas-phase flow in bulk and in porous media, a system providing flowing laser-polarized gas will also enable new studies of granular media, where the NMR observable gas is capable of fluidizing a granular system. NMR is fast becoming a popular tool for studying granular systems due to its ability to non-invasively probe the three-dimensional structure of the opaque system during motion. However, all current experiments have focussed on observing $^1$H NMR signal from the granular particles themselves, rather than the surrounding gas [77-82]. In order to facilitate this wide area of gas-phase NMR study, we have built and begun testing a $^{129}$Xe polarization system to provide a continuous supply of laser-polarized xenon, currently deliverable in a constant, controlled flow mode. In its final configuration, the system will also provide gas in a repeated, batch-delivery mode, supplying multiple shots of polarized xenon of a reproducible pressure and volume. In this paper, we report initial results from this gas delivery system in the constant flow mode, and investigate the parameter-space for studies of gas velocity by the NMR Fourier-encoding velocity method [12,20].

## MATERIALS AND METHODS

*Theory*

It is well known that the NMR echo signal observed in a PGSE experiment has a Fourier relationship to the probability of spin motion – the so-called displacement propagator, which can be thought of as a spectrum of motion. The echo signal, $E$, obtained in a PGSE experiment can thus be written as [12]:

$$E(\mathbf{q},t) = \int \overline{P}_s(\mathbf{R},t) \times \exp[i2\pi \mathbf{q} \cdot (\mathbf{R})]d\mathbf{R} \quad (1)$$

where $\overline{P}_s(\mathbf{R}, t)$ is the ensemble average displacement propagator, or the probability of a spin having a displacement $\mathbf{R} = \mathbf{r'} - \mathbf{r}$ proceeding from any initial position $\mathbf{r}$ to a final position $\mathbf{r'}$ during the 'diffusion time' $t$ (often referred to as $\Delta$ in the literature). $\mathbf{q}$ is the wavevector of the magnetization modulation induced in the spins by a field gradient pulse of strength $g$ and pulse duration $\delta$. The magnitude of $\mathbf{q}$ is $\gamma \delta g/2\pi$, where $\gamma$ is the spin gyromagnetic ratio. As a result, the Fourier transform of $E$ with respect to $\mathbf{q}$ yields an image of $\overline{P}_s$. In the limit of small $\mathbf{q}$, $\overline{P}_s$ is a Gaussian, and it can be shown that the spins undergoing motion will produce an echo with a phase factor $\exp[i2\pi qv\,t]$ where $v$ is the velocity of the spins, while the echo will be attenuated by a factor $\exp(4\pi^2 q^2 D(t)t)$ [12], where $D(t) = \langle [\mathbf{r'} - \mathbf{r}]^2 \rangle / 6t$ is the effective or time-dependent diffusion coefficient describing incoherent random motion of the spins in the pore space. $D(t)$, which describes the spin's mean-squared displacement, will decrease with increasing $t$, with more spins encountering barriers to their motion as $t$ is increased.

For small q measurements, at short diffusion times (i.e., small $t$), the fraction of fluid spins whose motion is restricted by pore boundaries is $\sim (S/V_p)\sqrt{D_0 t}$, where $S/V_p$ is the pore surface-area-to-volume-ratio and $\sqrt{D_0 t}$ is the characteristic free-spin-diffusion length scale for a diffusion time $t$. Using this basic concept, Mitra et al. have shown that for small $t$ [13,14]:

$$\frac{D(t)}{D_0} = 1 - \frac{4}{9\sqrt{\pi}} \frac{S}{V_p} \sqrt{D_0 t} + O(D_0 t) \qquad (2)$$

This important relation has been verified experimentally with NMR of liquids imbibed in a variety of model porous media [15], with the knowledge that for random bead packs, $S/V_p = 6(1-\phi)/(\phi b)$ [footnote 1] where $\phi$ is the sample's porosity and $b$ the bead diameter. If the effects of enhanced relaxation at pore boundaries can be neglected, then for very long $t$, $D(t)/D_0$ approaches an asymptotic limit [16]:

$$\frac{D(t)}{D_0} = \frac{1}{\alpha} \qquad (3)$$

where $\alpha$ is the tortuosity of the porous medium. In dense random beads packs $1/\alpha \sim \sqrt{\phi}$ [16]. This limit has been observed in bead packs and rocks only from gas-phase time-dependent diffusion coefficient measurements [55,56]. The Pade′ approximation has been successfully used in the past to extrapolate between these two limits for both liquid [15] and gas-phase $D(t)/D_0$ data [56,57]:

$$\frac{D(t)}{D_0} = 1 - (1 - \frac{1}{\alpha}) \times \frac{(4\sqrt{D_0 t}/9\sqrt{\pi})(S/V_p) + (1 - 1/\alpha)(D_0 t/D_0 \theta)}{(1 - 1/\alpha) + (4\sqrt{D_0 t}/9\sqrt{\pi})(S/V_p) + (1 - 1/\alpha)(D_0 t/D_0 \theta)} \qquad (4)$$

where $\theta$ is an adjustable fitting parameter which has units of time, and is expected to scale with the bead size [15]. More generally, $\sqrt{D_0 \theta}$ can be thought of a structural length scale, which for beads has been shown to be a constant fraction of the bead diameter [57].

Outside the small $q$ limit, $\overline{P}_s$ is no longer Gaussian. As such, the signal attenuation curve ($q$-plot), is no longer linear with $g^2$, even for single-component diffusion. When $t$ is large enough for spins to completely sample the restrictions of a given pore, $\overline{P}_s$ reduces to the pore spin density function, $\rho(\mathbf{r'})$, and the echo attenuation becomes the Fourier power spectrum of $\rho(\mathbf{r'})$ [10,11]. Specifically, for spins restricted in an open-pore system such as bead packs, the $q$-plot will exhibit a sinc modulation, with a minimum before rising to a maximum at a value of $q$ that corresponds to the reciprocal of the bead diameter. This phenomenon has come to be termed "NMR diffusive diffraction" or "NMR scattering". Other effects are manifested in one-dimensional restricted systems, and when flow is present [83].

Velocity measurements are obtained from PGSE-style experiments directly from the displacement propagator. The complex signal from each spectrum (or pixel in the case of a two-dimensional image) is Fourier transformed with respect to $q$ to yield the average propagator for those spins. The phase factor accumulated from coherent flow during $t$ manifests itself as an offset of the propagator from zero, which yields the average velocity experienced by the spins in the sample (or pixel) during $t$. Using the Fourier encoding velocity method, after transformation, the velocity is calculated from [12]:

$$v = (2\pi g_{steps} k_v) / (q_{fft} \gamma \delta t g) \qquad (5)$$

where $g_{steps}$ = the number of gradient values used (including zero), $k_v$ = the offset in digital points from zero of the maximum of the velocity spectrum, and $q_{fft}$ is the number of points Fourier transformed (including those used for zero-filling) with respect to $q$. In two-dimensional imaging, similar analysis for each pixel yields an image where the signal intensity in each pixel is indicative of this average velocity. Similarly, maps of the time-dependent diffusion coefficient can be produced by fitting the natural log of the echo attenuation, $\ln(E(q)/E(0))$ to the function $(4\pi^2 q^2 D(t)t) = (-\gamma^2 g^2 \delta^2 D(t)(t - \delta/3))$, i.e., the Stejskal-Tanner equation [2], yielding $D(t)$ on a pixel-by-pixel basis.

*Samples*

Samples of random packed spherical glass beads of different sample diameters were prepared for thermally polarized xenon experiments. Cylindrical glass cells of volume ~ 50 cm$^3$ held the bead-packs. Each cell contained beads of two different sizes, with beads of either 3 or 4 mm diameter mixed with beads of either 0.5 or 1 mm diameter. Appropriate amounts of xenon (isotopically enriched to 90% $^{129}$Xe) and oxygen were frozen inside the cells at liquid nitrogen temperature. The cells were then sealed and warmed to room temperature, allowing the gases to evaporate and establish the desired partial pressures inside the cell. Typically, this was ~ 5 - 6 bar xenon and ~ 1.5 bar oxygen.

Laser polarization was achieved by spin exchange collisions between the $^{129}$Xe atoms and optically pumped rubidium vapor. The gas mixture used for optical pumping contained either 92% xenon and 8% nitrogen, or 5% xenon, 10% nitrogen and 85% helium. The sample cell was optically pumped at ~

110° C for approximately 15 minutes, in a field of ~ 13 G from a small Helmholtz coil pair, using ~ 15 watts of circularly polarized light at 795 nm from a fiber-coupled diode laser array [Optopower Corp, Tucson, AZ]. After optical pumping, the polarization chamber was opened to a previously evacuated sample in the NMR magnet, connected by very thin teflon tubing. The rear end of the sample was connected to a Mass Flow Controller [MKS Instruments, Methuen, MA] which had a capacity for regulating flows of 0 to 1000 $cm^3$/min. A vacuum pump completed the circuit. Under the influence of the pump, and moderated by the Mass Flow Controller, xenon would flow continuously from the supply bottle through the polarization chamber, then to the sample in the NMR magnet, and finally to the pump. With a suitable supply of xenon gas mixture, stable flows could be maintained for many hours; however optimum polarization in the continuous flow mode was obtained for flow rates below 400 $cm^3$/min. The samples used for the experiments reported here included a 1.25 inch inner diameter plastic tube, 0.125 inch inner diameter flexible teflon tubing which was looped through the RF coil 6 times, and a glass cell that contained 2 mm glass beads.

*NMR technique*

In practice, the very short spin coherence time ($T_2$) of fluid spins in a porous sample, and the high background gradients that result from susceptibility contrast between the solid grains and the imbibed fluid [84] make the simple spin echo technique unsuitable for measuring $D(t)$ or for NMR scattering experiments. Instead, for measurements in porous systems, we used a modified stimulated echo sequence incorporating alternating bi-polar diffusion encoding gradient pulses (PGSTE-bp) which serve to cancel out the effect of the background gradients while applying the diffusion encoding gradient pulses [67,68]. The sequence is illustrated in Fig. 1a, where the timing parameters labeled correspond to the description in the previous section. The gradient pulses are half-sine shaped in order to reduce eddy current ring-down after the application of the gradient pulse. $D(t)$ was obtained from the resulting data by fitting the natural log of the measured echo attenuation to a modified form of the Stejskal-Tanner equation [68]:

$$\ln(S(g,t)/S(0,t)) = -g^2\gamma^2\delta^2(2/\pi)^2 D(t)(t - \delta/8 - T/6) \qquad (6)$$

where all timing parameters are described above, and the $(2/\pi)^2$ factor accounts for the half-sine shape of the gradient pulses. Low and high $q$ limitations of this method have been described elsewhere [56].

For experiments on laser-polarized gas flow in tubes without obstructions or glass beads, background gradient compensation was not necessary, and $T_2$ was long. Hence, the simpler PGSE-related methods were used, although modifications for use with laser-polarized gases were incorporated - specifically, the use of low flip angle excitation pulses and the removal of 180° pulses. For spectroscopy measurements, the Pulsed Gradient Echo was used [53], and imaging experiments were similarly performed using the spin-warp Gradient Recalled Echo technique with pulsed gradient echo (PGE-GRE) for diffusion/flow sensitization [58]. This is illustrated in Fig. 1b.

*NMR implementation*

For $D(t)$ measurements on the mixed bead packs with thermally-polarized xenon, the bi-polar stimulated echo technique (PGSTE-bp) was implemented on a GE Omega/CSI spectrometer (GE NMR Instruments, Fremont, CA), interfaced to a 4.7 T magnet. Experiments were performed at 55.348 MHz for $^{129}$Xe using a tuned home-made solenoid NMR RF coil. Applied magnetic field gradients up to 7 G/cm in strength were available. In these measurements, $\delta$ was fixed at 750 µs in all experiments, the minimum possible without distortion of the gradient shape on this system. $t$ was varied from a minimum of 25 ms to a maximum of 3 s. Values of $g$ were chosen to produce suitable attenuation of the $^{129}$Xe signal for measuring $D(t)$ [56]. The combined parameters resulted in $T = 4$ ms in all experiments. The inclusion of paramagnetic oxygen with the xenon reduced the $^{129}$Xe $T_1$ from tens of seconds to ~ 1.5 s, thereby enabling efficient signal averaging of the low NMR signal expected with thermally polarized $^{129}$Xe gas. Generally, 12 different $g$ values were used with a repetition rate of ~ 7-8 s, and the number of signal averaging scans ranged from 16 - 256, depending on sample line-width, signal strength and diffusion time $t$ used. All experiments were performed at room temperature.

All other experiments, which involved flowing laser-polarized xenon, were carried out using a Bruker AMX2 - based spectrometer (Bruker Instruments Inc., Billerica, MA) interfaced to a 4.7 T magnet. This system is equipped with a 12 cm ID gradient insert (Bruker) capable of delivering gradient pulses of up to 26 G/cm. We employed an Alderman-Grant-style RF coil, tuned to 55.35 MHz for $^{129}$Xe (Nova Medical Inc., Wakefield, MA). The PGSTE-bp method was implemented in a similar manner on this instrument for NMR flow scattering experiments with glass beads. Again, $\delta$ was fixed at 750 µs in all experiments. Values of $g$ were chosen to ensure $q$ exceeded the reciprocal of the bead size. $t$ was varied to reduce excessive diffusive signal attenuation while ensuring the spins could move further than one bead diameter during the diffusion time. The combined parameters resulted in $T = 4$ ms in all experiments. 16 different $g$ values were used and the repetition rate was varied with the flow rate to ensure enough time for freshly polarized xenon to flow completely into the sample chamber. Higher flow rates also permitted fewer signal averaging scans, as the laser-polarized xenon suffered less $T_1$ relaxation while travelling in the tube to the sample. Velocity measurements of unrestricted gas flow in tubes were obtained from images of a 5cm thick cross-sectional plane of the tubes. The image data were 64 × 32 matrices, zerofilled to 256 × 256 across a field of view of 100 mm, giving a final resolution of 390 μm². For velocity encoding, 8 different encoding gradients were applied, ranging from 0 to 17 G/cm. The gradient pulse time, $\delta$, was 1 ms, while the velocity-encode time, $t$, was 3 ms. As before, the repetition rate was varied with the flow rate, between 1 and 5 seconds. As a result, each velocity image, of 8 velocity and 32 phase encodes took between 2 and 40 minutes to acquire.

## RESULTS AND DISCUSSION

*Thermally-polarized Xe $D(t)/D_0$ in mixed bead samples*

Three separate static samples, each containing a mixture of two different sized glass beads and ~ 5 - 6 bar pressure of xenon gas, were used to obtain Xe $D(t)$ measurements in model heterogeneous (or multiple-length scale) porous media across a wide range of diffusion lengths. Data for two of these samples are shown in Fig. 2, plotted as the reduced diffusion coefficient, $D(t)/D_0$, as a function of the diffusion length, $\sqrt{D_0 t}$, where $D_0$ is the free gas diffusion coefficient. Data from the 3rd sample looked similar. When the samples were filled with gas, we estimated the porosity of the samples from the ratio of gas volumes required. Using the relationship for packed mono-sized spherical beads, $1/\alpha \sim \sqrt{\phi}$ [16], we determined the expected long time (tortuosity) asymptote for each sample to be $1/\alpha = 0.51$. In addition, $D(t)/D_0$ data from a mono-sized bead pack of each of the two beads making up each mixture is shown as an aid to the limits for the two sizes. This mono-sized bead data was first shown in [57].

For all three data series in each plot, the Padé approximation [Eq. (4)] was used to interpolate between the long and short $t$ ($S/V_p$) limits and hence show the expected trend of $D(t)/D_0$ in the intermediate $t$ region (i.e. for the mono-sized beads, we did not fit the data, but simply interpolated between the limits, so as to investigate the experimental data near these limits.) The Padé interpolations were made using the observed $1/\alpha = 0.64$ and 0.65 for 1 and 0.5 mm beads respectively. $S/V_p$ was calculated from the resultant porosity [$S/V_p = 6(1-\phi)/(\phi b)$ and $1/\alpha \sim \sqrt{\phi}$], giving $8.42/b$ and $8.22/b$ respectively (where $\phi$ is the porosity and $b$ is the bead diameter). For the 4 mm beads, the $S/V_p$ limit of $9.79/b$ was observed directly, and the expected tortuosity limit of $1/\alpha = 0.62$ was determined from the porosity. Deviations of the 4 mm bead results from the Pade′ interpolation at intermediate and long times is due to finite sample size effects [57]. Conversely, the deviation of the 0.5 mm bead data from the Pade′ line at short times results from diffusion during the diffusion-encoding gradient pulse that is significant in relation to the pore size. This results in incomplete background gradient compensation by the PGSTE-bp sequence, and violations of the Narrow Pulse Approximation, which voids the propagator formalism and the Stejskal-Tanner analysis method for PGSE experiments [56,57]. For the mixed bead sample, we expect the small sized beads to dominate the observed $S/V_p$ value. Hence, we fit the Pade′ line for the mixed bead samples using the $S/V_p$ value for the mono-sized 1 or 0.5 mm bead pack, in combination with the tortuosity limit of $1/\alpha = 0.51$ obtained from the porosity determination.

The $D(t)/D_0$ data from both mixed bead packs exhibit an unusual shape in comparison to the mono-sized bead samples. At the shortest possible diffusion times feasible, the data does indeed approach that of the smaller mono-sized bead pack (1 and 0.5 mm respectively). However, the $D(t)/D_0$ data then continues to decrease over a long distance that is more indicative of the behavior seen in 4 mm beads. While in this model system, the $D(t)/D_0$ decrease still appears monotonic, the shape of the data, which exhibits a very lengthy, slow decrease towards the $D(t)/D_0$ limit, is very similar to our original $^{129}$Xe data from a sample of Indiana Limestone rock [55].

A few simple relationships to characterize the mixed bead samples appear clear. While a maximum observable diffusion distance of ~ 1.8 mm ($t = 4$ s) is not quite long enough to reach the tortuosity asymptote, it appears the data is approaching the $1/\alpha = 0.51$ value. This verifies that the $1/\alpha \sim \sqrt{\phi}$ relationship, applicable in mono-sized bead packs, is also valid for a sample of mixed spherical beads. However, the relationship between porosity and $S/V_p$ is no longer applicable. In addition, as expected, the $S/V_p$ of the mixed bead system reflects that of the smaller bead in each pack. However, to fully probe this regime, further experiments should be carried out with water as the diffusing spin. We have shown [57] that for very small beads where the gas diffusion inside the pore during the application of the gradient pulse is significant on the pore scale, $S/V_p$ can be accurately determined using water. With $D_0 \sim 3$ orders of magnitude lower than gases, such short-$t$ effects are negligible in water diffusion. (Of course, this low $D_0$ makes long-length scale diffusion measurements impossible with water.) These short-$t$ diffusion effects result in the 0.5 mm bead data lying above the Pade′ line at short times. It is likely that an accurate determination of $S/V_p$ in the mixed bead pack, would see the Pade′ line for the mixed bead sample approach that of the theoretical line for a 0.5 mm pack, and not cross above it.

Given the current data, the Padé line can be made to fit the experimental data in the intermediate-$t$ regime very well. However, the Padé fitting parameter, $\theta$, and the related Padé length, $\sqrt{D_0\theta}$, are significantly different from values obtained in mono-sized bead packs. The Padé length was found to be ~ $0.13b$ across a wide range of bead sizes in mono-sized bead packs [57]. In the mixed bead packs, the fitted Padé lines in Fig. 2 yielded $\sqrt{D_0\theta}$ of $0.07b_{large}$ and $0.28b_{large}$ for the 4 & 1 mm and 4 & 0.5 mm bead packs respectively, where $b_{large}$ is the diameter of the larger bead in the pack. In addition, $\sqrt{D_0\theta} = 0.15b_{large}$ for a 3 & 0.5 mm bead pack, data which is not shown here.

The large diffusion lengths observed in these samples illustrate the power of this technique for studying heterogeneous porous structures. $^{129}$Xe atoms can diffuse ~ 2 mm or more before spin depolarization limits the measurement, dimensions similar to the long distance sample heterogeneity encountered in natural porous media, thus permitting an accurate measurement of tortuosity as well as the length-scale at which the sample becomes spatially homogeneous. Such multi-pore diffusion cannot be measured by NMR of liquids imbibed in porous media with pores > 50 μm [15,17]. Our initial results indicate that mixed-sized spherical bead packs can prove a useful model for studying multiple, longer length scales in porous media. Continued further investigation of $D(t)/D_0$ in mixed-sized spherical bead packs, however, is required, in order to fully understand the meaning of the macroscopic length scales that can be determined in such a heterogeneous system, and hence in natural samples such as reservoir rocks. It should be noted that, unlike the experiments with the mono-sized bead packs in ref [57], the xenon $D_0$ could not be measured in-situ in the sample containers used for these current experiments. Instead, $D_0$ was estimated based on the gas content transferred into each sample container, hence introducing a small uncertainty in the value of xenon $D_0$ used in the analysis. In addition, $S/V_p$ should be determined accurately using water diffusion measurements in each pack either immediately before or after the xenon measurements. Finally, careful attention to the proportion of each type of bead in the sample will permit computer simulations of diffusion in these samples, as an aid to understanding the information relating to macroscopic length scales that could be inferred from experimental data.

*Flowing laser-polarized xenon gas*
Initial tests of the continuous flow laser-polarized xenon facility involved measurements of the gas velocity at different flow rates while experiencing unobstructed flow in straight tubes or pipes. This allowed testing of the apparatus, as well as ensuring the effectiveness of velocity and diffusion mapping on flowing gas. Fig. 3 shows a velocity image of laser-polarized xenon flowing in a 1.25 inch ID tube. The image shows the gas flow velocity to be an almost constant and uniform value of 7 mm/s across tube, indicating a plug flow. This differs from many fluid flows, especially liquid flows, which exhibit a laminar flow profile in such conditions at moderate flow rates. The only prior NMR measurements of gas flow profiles indicate the possibility of obtaining laminar profiles in some cases in narrower tubes [60], but other examples of plug gas flow have also been observed [61]. It should be noted that this image was acquired using the 92% xenon gas mixture, and has a velocity resolution of 0.74 mm/s (obtained by evaluating Eq. 5 when $k_v = 1$). Higher flow rates were difficult to obtain in this experimental configuration. The large volume of the tube required such significant quantities of xenon that if the gas flowed faster, the xenon would move through the polarization chamber so quickly that it would not be laser-polarized to a sufficient extent to obtain a usable NMR signal.

In order to increase the gas velocity while maintaining similar mass flow rates, we conducted further experiments of gas flow in narrow teflon tubing with a 1/8 inch (~ 3 mm) internal diameter. The tubing was looped back and forth through the RF coil 6 times, giving flow in opposite directions in each of three tubes. Imaging was accomplished using the same method as for the large tube. Figs. 4a and 4b show velocity images of laser-polarized xenon flowing through this combination of tubes at mass flow rates of 100 and 200 $cm^3$/min. The velocity images clearly show the tubes containing xenon flowing in opposite directions, and that these velocities increase with the increase in flow rate. Something close to plug flow is observed, with a high average velocity in each tube. However, there is a slight variation in this average velocity at 100 $cm^3$/min, which becomes clearly obvious at 200 $cm^3$/min, as much as ~ 25% in some tubes. Some variation may be expected as the tubing was bent tightly outside the RF coil, and plug flow with a uniform velocity may not have been re-established in the distance until the tube was again in the active region of the coil. In addition, no apparent velocity gradient is seen at the tube wall. While gas flow will exhibit much lower wall shear interactions than liquid flow, it may be that improved imaging resolution (true resolution in this experiment, before zero filling, exceeded 1 $mm^2$) would indicate a sharp velocity gradient at the tube wall, as has been seen previously [61].

The tubes do not appear perfectly circular in the image, as the tubing is flexible and bends through the plane of the image slice. In particular, two tubes close together (at lower right) cross one another within the image plane, resulting in the velocity image for one tube showing mostly negative flow, while one side of the tube appears to exhibit positive flow. Below the images, at bottom left, is an example velocity spectrum (displacement propagator) from one pixel in Fig. 4b. This propagator is typical of those in nearly all pixels in the image, showing the displacement distribution to be Gaussian, indicating coherent, uniform flow for that pixel (but not illustrating the variation observed from one pixel to the next). The one exception to this rule is the propagators from some pixels in the lower right tube, which exhibit two Gaussians, a positive one having slightly higher probability than the negative. As such the velocity image for this tube appears positive in this region, but the propagator indicates that two different coherent flows, in opposite directions, are occurring within these pixels within the image slice. Were the flow in this region of the tube turbulent and changing direction during the experiment, the propagator would lose its Gaussian shape, and indicate that no single velocity has a high probability of being observed.

Effective diffusion coefficient maps for the xenon flowing in these narrow tubes are shown in Figs. 4c and 4d, and are obtained from the same datasets that yielded the velocity images in Figs 4a and 4b respectively. We note the diffusion coefficient observed in all tubes is independent of the direction of the flow of xenon, confirming that random Brownian motion occurs independent of any imposed flow field, and that the PGSE method can distinguish between these two motions [12]. Apart from some minor edge effects and the artifact region in the lower right tube, the observed diffusion coefficient generally appears uniform in each tube, especially at the lower flow rate. However, there is a clear increase observed as the gas flow rate increases, changing the observed diffusion coefficient from ~ $7 \times 10^{-6}$ to ~ $10 \times 10^{-6}$ m$^2$/s as the flow rate increases from 100 to 200 cm$^3$/min. This increase does not reflect an increase in true self diffusion, but is due to Taylor dispersion effects. As spins cross from streamlines of one flow to another, especially at 200 cm$^3$/min, where there is greater variability in the flow within a tube, their apparent diffusion is enhanced [83,85]. In addition, greater variation in the apparent diffusion coefficient is observed at this flow rate. When using a single gradient pulse pair, the PGSE experiment remains susceptible to such dispersion effects, and at very high flow rates, the Stejskal-Tanner method yields the dispersion, rather than diffusion, coefficient. Implementation of a double-PGSE method removes imposed coherent flow effects on the diffusion method, and can still yield the true self-diffusion coefficient in such instances [83].

The images in Fig. 4 show the efficacy of the PGSE techniques, combined with imaging, to study laser-polarized xenon flow using NMR at high flow rates; velocities of up to 180 mm/s were recorded at higher flow rates. However, velocity resolution is an important factor to consider when implementing this technique to study gas flow. For these images, the 5 % xenon mixture was in use. The reduced xenon partial pressure (replaced by light gases $N_2$ and He) resulted in a significantly higher xenon diffusion coefficient in this mixture, which served to reduce both the minimum feasible imaging resolution and velocity resolution. For these images, the minimum velocity resolution was ~ 3.5 mm/s, about a factor of 5 higher than for the image in Fig. 3 where a 92% xenon mixture was used. By contrast, these values are both considerably higher than the velocity resolution obtainable when studying liquid flow, which can be as low as 10 μm/s [12,21]. This is apparent from Eq. 5, from which the velocity is calculated from the offset from zero of the displacement propagator. Essentially all terms in this equation relate to the timing and strength of the flow-encoding gradients, outlining the gradient strength required to create a sufficient magnetization phase grating such that any phase offset due to coherent flow can be observed. Therefore similar gradient strength/time is required to observe a given flow rate in liquid or gas, and there is little room for the consideration that significant gradient strength can completely attenuate the gas-phase signal due to the much higher gaseous diffusion coefficient. Therefore, diffusive attenuation will generally limit gas flow velocity resolution to ~ 0.5 - 1 mm/s, in much the same way that gas-phase imaging resolution is limited to ~ 0.5 - 1 mm [61,86]. To some degree, this limitation is a minor concern, as gas flows of 10 - 100 mm/s, especially in porous media, are much more easily obtainable than with liquid flow, where the pressures required to maintain such flow rates would be prohibitive. Previous studies of gas flow measurement by NMR have addressed the imaging resolution issue, but not the limitation to velocity resolution, as gas flow rates of 20 - 1000 mm/s were being observed with no attempt to measure slower flows [58,60,61]. By contrast, there is no limitation to measuring gaseous diffusion coefficients, orders of magnitude higher than those of liquids, using the Stejskal-Tanner method, provided sufficient resolution of extremely small pulsed gradient increments of 0.1 G/cm or less can be maintained by the spectrometer.

A final test of the continuous flow laser-polarized xenon apparatus involved flowing the gas through a 2 mm glass bead pack, and studying the echo attenuation in a PGSTE-bp spectroscopy experiment. The echo attenuation plots are given in Fig. 5, and show the initial monotonic decrease at low flow rates giving way to a sinc attenuation profile, related to the bead pack structure, at the highest flow rates. For a system of uniform packed beads with diameter $b$, diffusion over time scales long enough for spins to experience the full pore heterogeneity will result in a "NMR scattering" or "diffusive diffraction" pattern, where the echo attenuation maximum that follows the first minimum occurs at $q = b^{-1}$ [10,83]. It can be easily shown that the first minimum occurs at $q \approx 0.7 b^{-1}$. This pattern can be enhanced by flowing the spins, essentially helping them to fully sample the pore heterogeneity faster than they otherwise would by diffusion alone [23,83]. The use of gases to obtain such a pattern for the first time indicates that gas-phase NMR can be used to probe regular structures of a much larger length scale, on the order of mm, than can be probed by liquid-phase NMR scattering measurements (10 - 100 µm). This is true even when the NMR observable spins are under the influence of flow - as considerably higher gas flow rates can be obtained with gases than with liquids.

The current experiments show a distinct scattering minimum at ~ 323 $m^{-1}$, corresponding to an average bead size of 2.17 mm, close to the nominal value of 2 mm. Because of significant gas hold-up within the bead pack, the majority of the spins move much slower than the average velocity based on the mass flow rate. As a result, even at the flow rates obtainable with the current system, many spins are traversing less than 2 mm in the measurement time (500 ms) and as such, the maximum at $q = b^{-1}$ is not clearly defined. Nonetheless, the advantage of gas-phase NMR scattering for probing regular pore structures on the mm scale is obvious. As with velocity resolution, the NMR scattering wavevector also makes no allowance for the higher gas diffusion coefficients, requiring almost the same gradient strength/time to probe a given length-scale, regardless of the nucleus being observed. However, because of the reciprocal relationship of $q$ to $b$, when probing structures on the mm length scale, as can only be done with gas at high flow rates, less gradient strength is required to probe for relevant structural factors. This makes gas-phase NMR scattering a complementary technique to liquid NMR scattering, based on the length scale to be probed.

## CONCLUSIONS

Gas-phase NMR diffusion and flow measurements are a powerful tool for the study of porous media, granular systems, and gas-flow dynamics. In the current work, we have used thermally polarized xenon imbibed in random packs of mixed-sized glass beads to study the approach of the effective or time-dependent diffusion coefficient to the long-time limit, with the aim of understanding the determination of long length scale structural information in a heterogeneous system. We have also carried out initial tests of continuous flow laser-polarized xenon, both unrestricted through tubing of two different diameters, and through a 2 mm bead pack.

The large gas diffusion coefficients, ~ 3 orders of magnitude higher than liquids, permit the probing of long length scales in porous media, and provide the only manner of determining the heterogeneous length scale in porous media - by observing the approach of $D(t)/D_0$ to the tortuosity asymptote. We found that $D(t)/D_0$ of the xenon gas in mixed size bead packs at short diffusion times followed the same decrease as a pack of the smaller sized beads alone, hence reflecting the pore surface-area-to-volume-ratio of such a sample. The approach to the long-time limit follows that of a pack of the larger sized beads alone, although this limiting $D(t)$ is lower, reflecting the lower porosity of the sample compared to that of a pack of mono-sized glass beads. We obtained an excellent fit to the experimental data between the short and long diffusion time limits using the Pade′ approximation. However, the adjustable fitting parameter, the Pade′ length, $\sqrt{D_0\theta}$, did not yield a constant value or constant fraction of either bead size for these samples, as had been observed for mono-sized bead packs. Future work with these samples will include measuring $D_0$ in-situ rather than relying on a calculation based on intended gas partial pressures when filling the samples, and to use water $D(t)/D_0$ as a complementary method to unambiguously probe the pore surface-area-to-volume ratio of these samples. This will improve our ability to relate $\sqrt{D_0\theta}$ to structural length scales in heterogeneous systems

Initial studies of continuous flow laser-polarized xenon show we have the ability to image high velocity gas flows (20 - 200 mm/s), permitting the study of fluid flow rates much higher than can generally be obtained with liquids. We measured a plug flow profile of xenon gas in a 1.25 inch diameter tube (7 mm/s velocity) and near-plug flow in multiple loops of narrow 1/8 inch diameter tubing (20 - 180 mm/s). From the same data, we also determined apparent xenon gas diffusion coefficients, which were unaffected by the direction of the flow, but were seen to steadily increase with gas flow rate due to the effects of Taylor dispersion. We also presented the first gas-phase NMR scattering, or diffusive-diffraction, data. Flow enhanced structural features were observed in the echo attenuation data from laser-polarized xenon flowing through a 2 mm glass bead pack when flow rates exceeded 400 cm$^3$/min, the scattering minimum occurring at $q$ ~ 323 m$^{-1}$, which indicates a bead diameter of ~ 2.17 mm.

Gas velocity imaging has the potential to be a significant tool in porous and granular media study, as well as in gas fluid dynamics. The much higher gas flow rates permissible allow NMR scattering techniques to probe much longer regular length scales than could be probed with liquids. The ability to observe a NMR sensitive gas with high SNR will also allow future studies of the gas dynamics in granular media, especially gas-fluidized systems. While powerful, we note that gas velocity imaging has a diffusive-attenuation-induced velocity resolution limit of about 1 mm/s, in a similar manner to the diffusive-limited gas imaging resolution limit in MRI applications.

## FOOTNOTE (to page 8)

This relation is derived from the basic equations for surface area and volume of a sphere. Assuming $r$ = sphere radius and $b$ = sphere diameter:

$$\frac{S_{sphere}}{V_{sphere}} = \frac{4\pi r^2}{4\pi r^3/3} = \frac{3}{r} = \frac{3}{b/2} = \frac{6}{b} \tag{F1}$$

Furthermore, in a porous solid with total volume $V_T$, and porosity $\phi$, the volume of the pore space is $\phi V_T$ and the volume of the solid $(1-\phi)V_T$. Therefore, for a porous solid made up of packed spheres, the pore surface are to volume ratio $(S/V_p)$ is given:

$$\frac{S_{pore}}{V_{pore}} = \frac{S_{sphere}}{V_{pore}} = \frac{S_{sphere}}{V_{sphere}} \frac{V_{sphere}}{V_{pore}} = \frac{6}{b} \frac{(1-\phi)V_T}{\phi V_T} = \frac{6(1-\phi)}{\phi b} \tag{F2}$$

## ACKNOWLEDGEMENTS

We acknowledge Dr. Sam Patz (Brigham and Women's Hospital) for access to the 4.7 T instrument on which the mixed bead $D(t)$ data was acquired, and Dr. Pabitra Sen (Schlumberger-Doll Research) for useful discussions. This work was supported by the NSF, NASA, and the Smithsonian Institution.

# FIGURE CAPTIONS

Figure 1: (a) Pulse sequence diagram for the Pulsed Gradient Stimulated Echo (PGSTE-bp) with alternating bi-polar gradient pulses, as used in this work. The diffusion encoding gradient pulses, of length $\delta$ and strength $g$, are shown in gray, while crusher gradients are shown in black. The diffusion time is denoted $t$ and the total diffusion encoding time is $T$. See text for further description. (b) Pulsed Gradient Echo with Gradient Recalled Echo sequences for dynamic MRI. *g*, *gsl*, *gpe* and *gro* refer to the magnitude of the velocity encode, slice select, 2D phase encode and readout gradients, respectively.

Figure 2: Time-dependent diffusion measurements for thermally polarized xenon gas at ~ 5 - 6 bar pressure imbibed in randomly packed spherical glass bead samples. The series in triangles shows data from samples with two different sized beads: (a) 4 and 1 mm beads, (b) 4 and 0.5 mm beads. The series in squares shows data from two samples containing mono-sized beads of the same diameter as those in the corresponding mixed bead pack. The $^{129}$Xe time-dependent diffusion coefficient is normalized to the free gas diffusion coefficient, $D_0$, and plotted versus the diffusion length $\sqrt{D_0 t}$ in mm. The calculated long-*t* (tortuosity) limit for the mixed bead pack is shown as a solid straight line. Pade′ lines interpolating between the short-*t* ($S/V_p$) and long-*t* (tortuosity) limits for each of the 3 data series are also shown in solid, dash and dash-dot lines.

Figure 3: Velocity image of laser-polarized xenon flowing through a 1.25 inch diameter tube, obtained with the PGE-GRE method. The image intensity indicates a reasonably uniform velocity of 7 mm/s across the tube. Velocity-sensitive parameters included: $\delta = 2$ ms, $t = 8$ ms, $g = 0 - 22$ G/cm in 16 increments. Image parameters: original field of view = 80 mm$^2$, matrix size 32 µ 32, slice thickness = 2 cm, TE/TR = 15.2/1024 ms, 1 signal averaging scan (NS = 1). Total acquisition time ~ 9 min.

Figure 4: Velocity and diffusion images of laser-polarized xenon flowing at two different flow rates through 1/8 inch diameter tubing looped back and forth through the RF coil 6 times. (a) Velocity image at 100 cm$^3$/min flow rate. (b) Velocity image at 200 cm$^3$/min. (c) Diffusion image at 100 cm$^3$/min. (d) Diffusion image at 200 cm$^3$/min. (e) Velocity spectrum (propagator) from the pixel marked *e* in image b). (e) Signal attenuation plot from the pixel marked *f* in image d). Velocity-sensitive parameters included: $\delta = 1$ ms, $t = 3$ ms, $g = 0 - 22$ G/cm (for 100 cm$^3$/min) or 17 G/cm (for 200 cm$^3$/min) in 8 increments. Image parameters: original field of view = 80 mm$^2$, matrix size 64 µ 32, slice thickness = 5 cm, TE/TR = 10.4/5000 ms, NS = 4. Total acquisition time ~ 21 min.

Figure 5: Echo attenuation plots, $\ln(S(q)/S(0))$ vs $q$ ($q= \gamma\delta g/2\pi$), from a PGSTE-bp spectroscopy experiment for laser-polarized xenon flowing through a pack of 2 mm glass beads. The initial monotonic decrease at low flow rates gives way to a sinc attenuation profile related to the bead pack structure at the highest flow rates. Motion-sensitive parameters included: $\delta = 0.75$ ms, $t = 500$ ms, $g =$ 0 - 3.8 G/cm in 32 increments. TR ranged from 20 to 12 seconds, and NS from 16 to 6 as flow rates increased, optimized each time for suitable NMR signal. Slower flow rates resulted in higher initial xenon polarization as the spins spent longer in the polarization chamber, but suffered greater $T_1$ relaxation in the teflon tubing before reaching the RF coil. Maximum signal was obtained for flow rates ~ 400 cm$^3$/min.

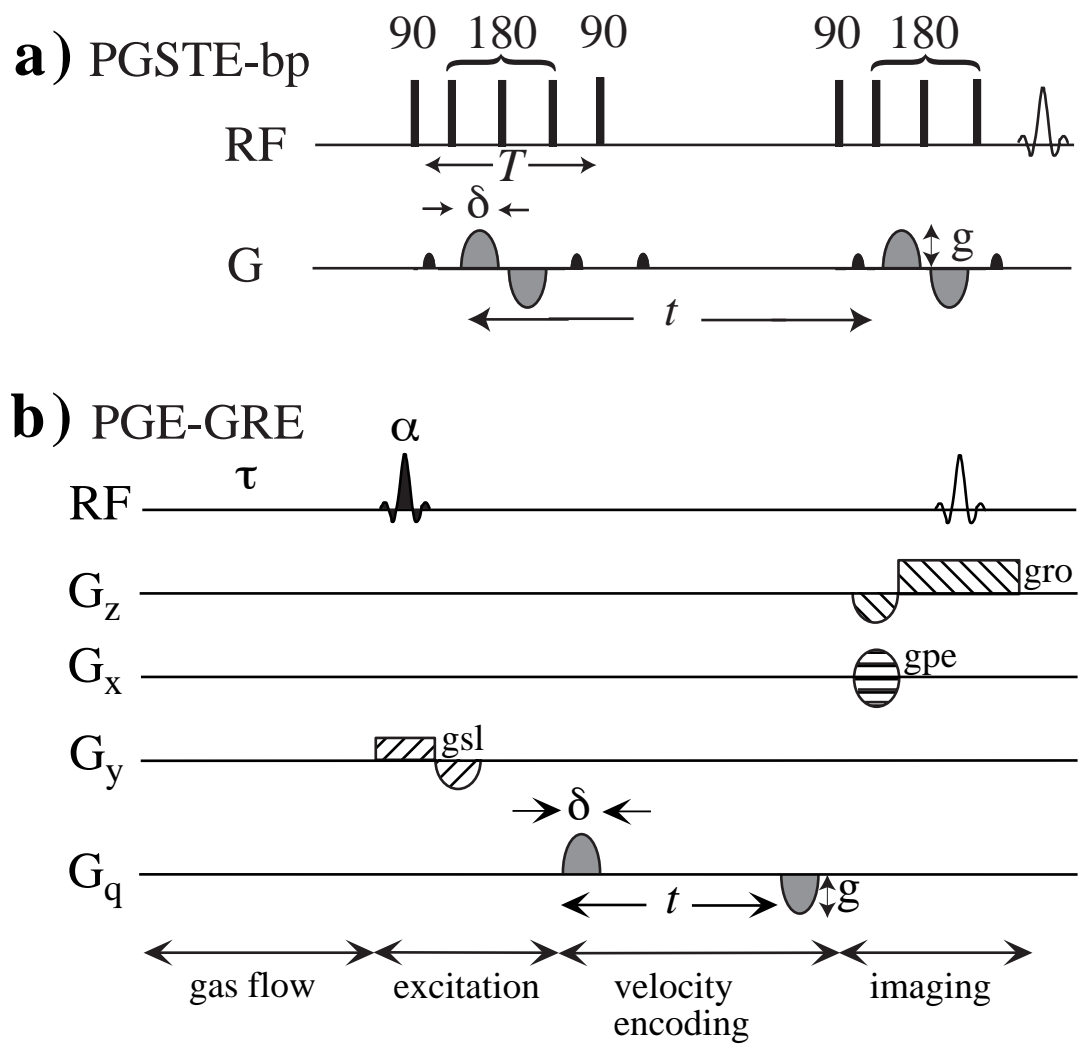

**Figure 1**

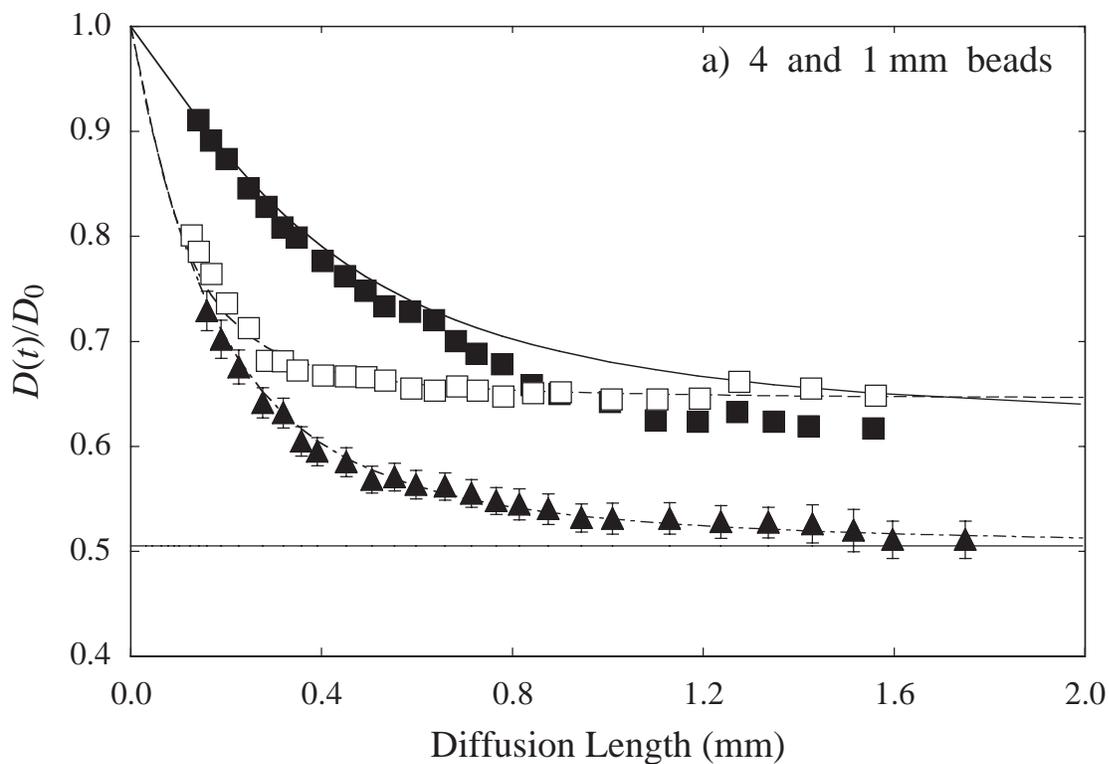

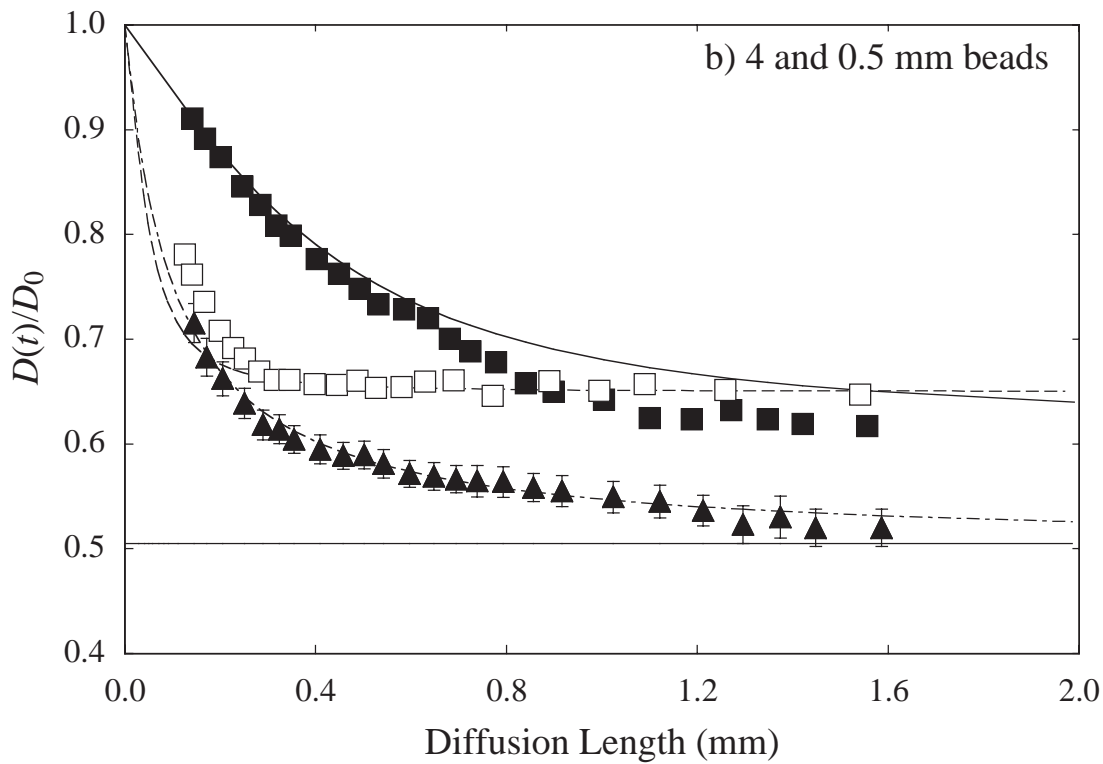

**Figure 2**

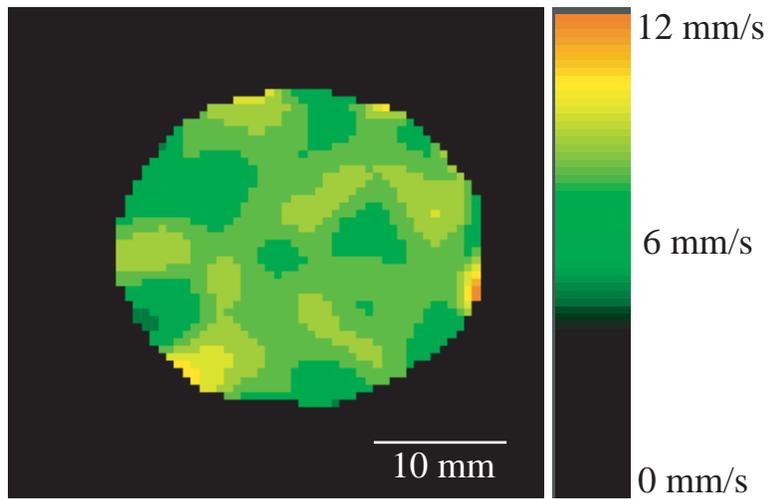

**Figure 3**

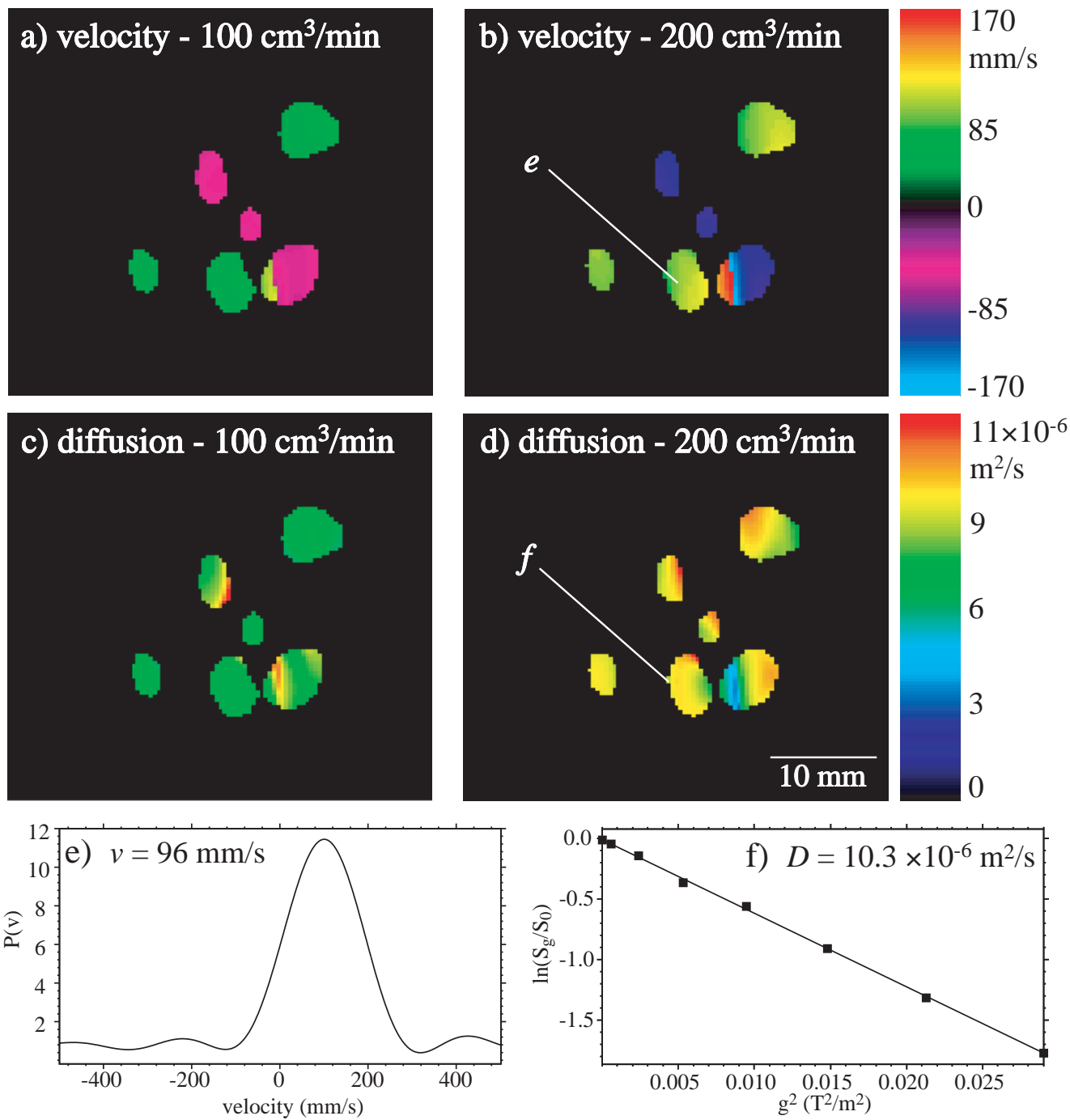

**Figure 4**

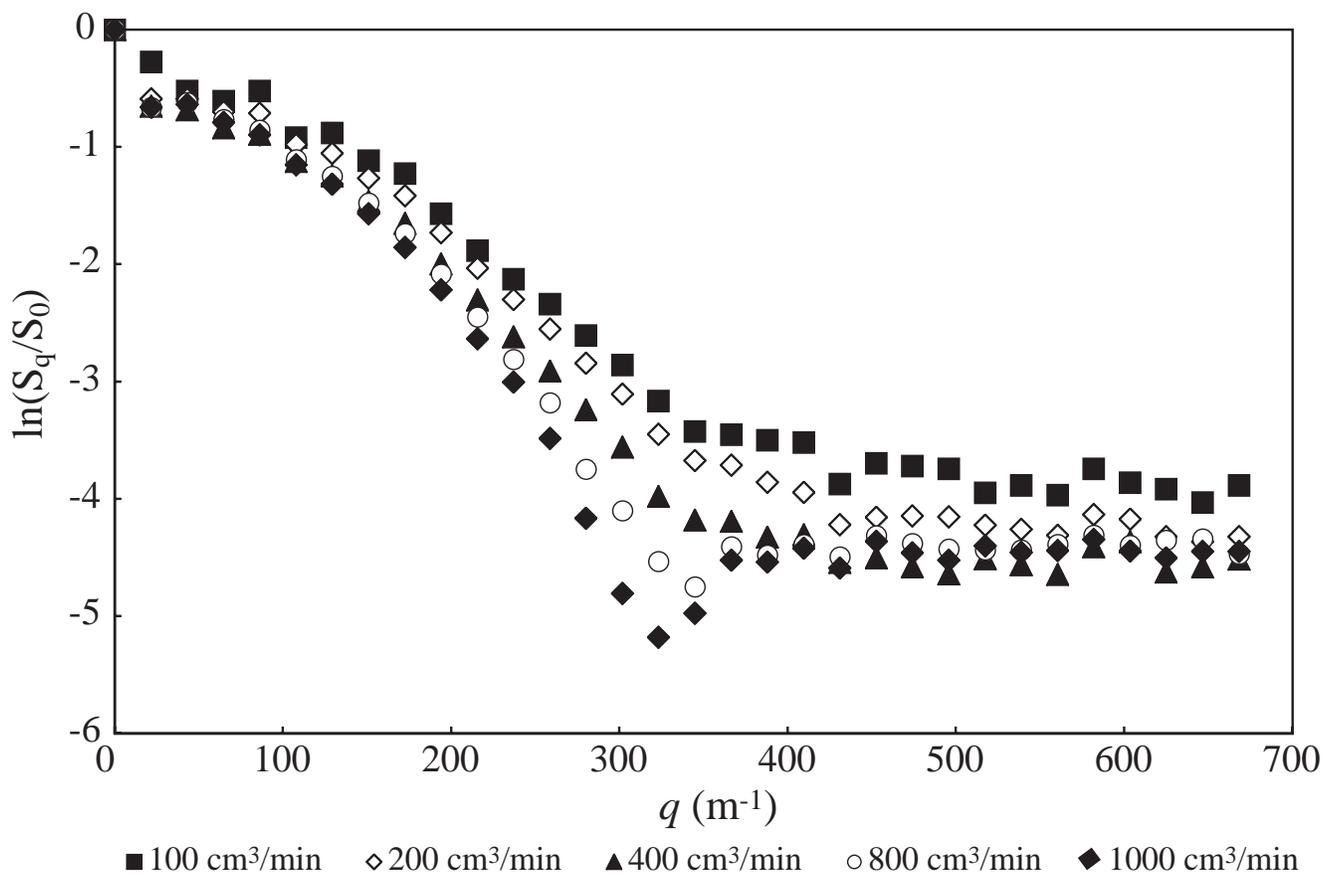

**Figure 5**